\documentstyle[psfig,aps,prl,amsfonts,amssymb]{revtex}
\begin{document}
\title{The Brody-Hughston Fisher Information
Metric}
\author{Paul B. Slater}
\address{ISBER, University of
California, Santa Barbara, CA 93106-2150\\
e-mail: slater@itp.ucsb.edu,	
FAX: (805) 893-7995}

\date{\today}

\draft
\maketitle
\vskip -0.1cm

\begin{abstract}
We study the
interrelationships between the Fisher information metric recently
introduced, on the basis of maximum entropy considerations,
 by Brody and Hughston (J. Math. Phys. 41, 2586 [2000])
 and the  monotone metrics, as explicated by Petz and Sud\'ar 
(J. Math. Phys. 37, 2662 [1996]).
 This {\it new} metric turns out to be {\it not}
strictly monotone in nature, and to yield --- {\it via} its 
normalized {\it volume element} --- a {\it prior} probability distribution
over the Bloch ball of two-level quantum systems 
that is {\it less} noninformative than those obtained from any of
the monotone metrics, even the {\it minimal} monotone (Bures) metric.
We best approximate the {\it additional}
 information contained in the Brody-Hughston prior over that
contained in the Bures prior by constructing a certain Bures
{\it posterior} probability distribution. This is proportional to the product
of the Bures prior and a likelihood function
based on {\it four} pairs of
spin measurements oriented along the diagonal axes of an inscribed cube.
\end{abstract}

\vspace{.1cm}

\hspace{1cm} {\bf{Key words: monotone metric, two-level quantum
systems, operator monotone function, Bures metric, 
Jeffreys' prior, Bloch sphere/ball,
noninformative prior, Bayesian analysis, Platonic solids, maximum entropy,
complex multivariate normal distribution}}

\vspace{.15cm}
\section{Introduction}
In one of their numerous recent contributions 
\cite{bh1,bh2,bh3,bh4} to the study of
the geometry of quantum mechanics, Brody and Hughston (BH) developed a
``method for representing probabilistic aspects of quantum systems by
means of a density function on the space of pure quantum states,'' a maximum
entropy argument allowing them ``to obtain a natural density function that
only reflects the information provided by the density matrix'' \cite{bh1}. 
(A similarly-motivated study --- based, in part, on extensive
work of Band and Park \cite{band1,band2} --- can be found in 
\cite{slater1}.) BH indicated how to associate a Fisher 
information metric
to their family of density functions 
({\it i. e.}, probability distributions),
each distribution, of course, corresponding to a density matrix.

In this study, we investigate, in the context of the two-level quantum
systems, the interrelationships of the BH
(Fisher information)
metric to the important, fundamental class of
(stochastically) 
monotone metrics \cite{ps,m1,m2,grasselli}. The monotone metrics are
{\it quantum} extensions of the (classically {\it unique})
 Fisher information metric. They fulfill the information-theoretic
desideratum of being nonincreasing under ``coarse-grainings''.
Let us note, in particular, that the
 much studied {\it Bures} metric 
\cite{b1,b2,b3,b4}, in fact, plays the role of 
the {\it minimal} monotone metric. A number of other 
(non-minimal) monotone 
metrics have been subjects of detailed investigation, as well
 \cite{mm1,mm2,mm3}. 

First, we find that while the {\it normal/tangential} component of the BH  
metric over the
``Bloch ball'' of two-level quantum systems appears to be consistent
with monotonicity, its {\it radial} component is 
conclusively {\it not} (cf. \cite{ozawa}).
We are able to verify this failure of monotonicity by finding an
example of a pair of density matrices,
the BH distance between which {\it increases} under a certain
completely positive trace-preserving mapping (coarse-graining).
 Also, as a {\it prior} distribution for Bayesian analyses,
the normalized {\it volume element} ($p_{BH}$) of the BH metric proves to be
considerably {\it less}
noninformative in nature than {\it any} of the normalized volume elements
of the monotone metrics 
(even including the Bures/{\it minimal} monotone metric). Following
the work in \cite{slaterclarke,bernie}, we
approximate the additional information so contained in $p_{BH}$ over that
contained in the prior ($p_{B}$) based on the Bures metric
by the construction of a certain Bures {\it posterior} distribution ($P_{B}$).
This is the normalized form of the product of $p_{B}$
and  a {\it likelihood} function based
on {\it four} pairs of  spin measurements oriented along the diagonals
of an inscribed cube.
Despite this noninformativity disparity, a certain strong congruence 
between the BH and Bures metrics emerges.
We also study, {\it en passant}, a number of other metrics of interest,
both monotone and non-monotone in character.
\section{Monotonicity analyses}
BH showed (by integration of a Gaussian distribution) 
that their generating function for the two-level quantum systems
could be written as
\cite[eq. (18)]{bh1},	
\begin{equation} \label{genfunct}
Z(\lambda)= (2 \pi)^3 {e^{-\lambda_{2}} - e^{-\lambda_{1}} \over 
\lambda_{1}- \lambda_{2}},
\end{equation}
where $\lambda_{1}$ and $\lambda_{2}$ are the eigenvalues of the
corresponding 
$2 \times 2$ density matrix.
The Hessian matrix of $\mbox{ln} Z(\lambda)$ 
\cite[eq. (16)]{bh1} then has the interpretation of
being a Fisher information matrix on the parameter space of the
(three-dimensional) family of BH probability distributions.

Choosing to parameterize the $2 \times 2$ density matrices ($\rho$) by
Cartesian coordinates ($x_{1},x_{2},x_{3}$) in the Bloch ball,
\begin{equation}
\rho={1 \over 2} \pmatrix{1+x_{1} & x_{2} +\mbox{i} x_{3}  \cr x_{2} 
- \mbox{i} x_{3}  & 1 - x_{1}  \cr}, \quad \Sigma_{i=1}^{3} x_{i}^{2} \leq 1
\end{equation}
we have that $\lambda_{1},\lambda_{2} 
 = {1 \pm \sqrt{\Sigma_{i=1}^3 x^{2}_{i}} \over 2}$,
so that the generating function (\ref{genfunct}) can be reexpressed as
\begin{equation}
Z(x_{1},x_{2},x_{3}) 
= 16 \pi^3 {\sinh{{1 \over 2} \sqrt{\Sigma_{i=1}^3 x^{2}_{i}}}
\over \sqrt{e} \sqrt{\Sigma_{i=1}^3 x^{2}_{i}}}.
\end{equation}
Then, computing the corresponding $3 \times 3$ 
Fisher information matrix,
$|| { \partial^2 \mbox{ln} 
Z(x_{1},x_{2},x_{3}) \over \partial  x_{i} \partial x_{j}}||$,
and converting to spherical coordinates
($x_{1} = r \cos{\theta}, x_{2} = r \sin{\theta} \cos{\phi},
x_{3} = r \sin{\theta} \sin{\phi}$), we obtain the 
(diagonal) BH Fisher information
metric for the two-level quantum systems,
\begin{equation} \label{BHmetric}
ds_{BH}^2 = {4 - r^2 \mbox{csch}^{2}({r \over 2})  
\over 4 r^2}  dr^2 + 
 {r \coth({r 
\over 2}) -2 \over 2 r^2} dn^2,
\end{equation}
where for the normal
 (as opposed to radial) component of the metric, 
$d n^2 = r^2 d \theta^2 + r^2 \sin^2{\theta} d \phi^2$.
Now, in spherical coordinates, a monotone metric on the 
three-dimensional space
of $2 \times 2$ density matrices must assume the form \cite[eq. (17)]{ps},
\begin{equation} \label{PS}
ds_{monotone}^2 ={1 \over 1-r^2} d r^2 +
{1 \over 1+r} g \Big( {1-r \over 1+r} \Big)  d n^2.
\end{equation}
(Unfortunately, no explicit demonstration of this proposition seems to
be available in the literature.)
Here, $g(t)= 1/f(t)$, where $f(t)$ is an {\it operator} monotone function
on the positive real axis
such that $f(t) = t f(t^{-1})$ for every $t>0$ \cite[Thm. 3.1]{ps}.
(A function $f$ is called operator monotone if for all pairs
of Hermitian operators satisfying $A \geq B$, that is $A-B$ is
positive semidefinite [all eigenvalues of $A-B$ being nonnegative], we have
$f(A) \geq f(B)$ \cite{bhatia}.)

So, since the radial component in the BH (Fisher information) metric 
(\ref{BHmetric}) is {\it not} identically
equal to ${1 \over 1 -r^2}$, we can conclude that
this BH metric does not strictly fulfill the role 
of  a monotone metric
over the two-level quantum systems in the sense of Petz and Sud\'ar
\cite{ps}. 
(Presumably, it is also {\it not} 
monotone for the higher-dimensional quantum
systems, but we do not at all address this question
here.) In Fig.~1 we display these two radial components. 
(In the vicinity of the fully mixed [classical] 
state, $r=0$,
the two radial components both behave approximately as
 {\it constants} (cf. \cite{slaterpre}), that is, 
1/12 in the BH case and 1 in the general monotone case.)
\begin{figure}
\centerline{\psfig{figure=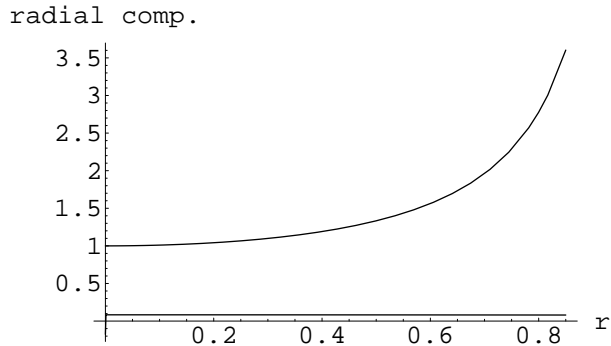}}
\caption{Radial components of an arbitrary monotone metric (\ref{PS}) --- the
upper curve --- and the  BH metric (\ref{BHmetric})}
\end{figure}

However, if we equate the {\it normal}  components of the two metrics
((\ref{BHmetric}), (\ref{PS})),
we obtain (setting 
first  $r = (1-t)/(1+t)$ and then taking $f(t) =1/g(t)$),
\begin{equation}  \label{monotone}
f(t)=-\left( \frac{{\left( -1 + t \right) }^2}
    {2\,\left( 1 + t \right)  + 
      \left( -1 + t \right) \,\coth (\frac{1 - t}{2 + 2\,t})}
    \right).
\end{equation}
A plot (Fig.~2) 
of $f(t)$, along with its first-order series expansion
about $t=0$,
\begin{equation} \label{series}
f(t) \approx { -3 + 4 e - e^2  + (7- 16 e + 5 e^2) t \over (e-3)^2} \approx
6.09929 + 5.70491 t = 12 (.508274 + .47541 t),
\end{equation}
reveals $f(t)$ to clearly be a monotonic
 (remarkably {\it linearly} increasing) function
on [0,1]. (Whether or not $f(t)$ also fulfills the criteria to be 
an  {\it operator} monotone function
remains to be formally determined, however.)
\begin{figure}
\centerline{\psfig{figure=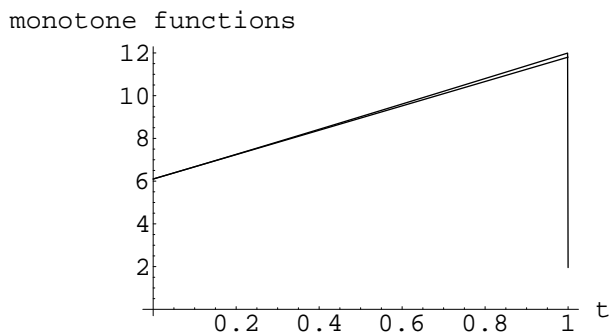}}
\caption{Monotone function (\ref{monotone}) imputed to the 
normal component of the BH Fisher information
metric (\ref{BHmetric}) on the two-level quantum systems, along with its
(slightly smaller) linear approximation (\ref{series}) 
 for $t \in [0,1]$}
\end{figure}
 If we were to divide $f(t)$
by 12 so it met the ``Fisher-adjustment'' requirement 
(satisfied by {\it all} monotone metrics) of having the value 1
at $t=1$,  it would then 
have the value .508274 at $t=0$, remarkably
close to the requirement of equalling .5 (met {\it only}
 by the Bures metric) 
for ``Fubini-Study-adjustment'' (on the pure states) 
\cite{zs}. (BH performed their integrations over the ``quantum 
phase space'' with respect to the Fubini-Study measure \cite{bh1}.) 
Also, for the Bures
 metric, $f(t) = .5 +. 5 t$, which rather closely
resembles $.508274 + .47541 t$, obtained {\it via} 
the series expansion (\ref{series}).

Consider an arbitrary pair of 
 $2 \times 2$ density matrices, $\rho_{1}$ and $\rho_{2}$.
Then for any completely positive 
trace-preserving map $\Phi$ on the space of operators
such that $||\Phi||<1$, we must 
have $d_{mon}(\Phi \rho_{1},\Phi \rho_{2}) \leq
d_{mon}(\rho_{1},\rho_{2})$, where $d_{mon}$
 is the distance function corresponding
to a monotone metric. We have, in fact,
 succeeded (through a random search process)
 in  constructing sets of 
$\Phi,\rho_{1},\rho_{2}$  for which this
inequality is {\it violated} (but quite rarely) for the BH metric.
(We verified our MATHEMATICA program by finding no violations when we
replaced the BH metric in the program
by the minimal or maximal monotone metrics.)
If we take Bloch coordinates, $r=.646675,\theta=2.51509,\phi=5.89259$
for $\rho_{1}$ and add the small differentials 
$4.17588  \cdot 10^{-6}, -8.44724 \cdot 10^{-6},
7.82807  \cdot 10^{-6}$  to
them, respectively, we generate $\rho_{2}$. Substituting these values into the formula
(\ref{BHmetric}) for $ds^2_{BH}$,
 we obtain a distance of $2.14985  \cdot 10^{-6}$.
 Now, setting
$u=2.43564, v=.0289153 $ in the trigonometric parameterization of $\Phi$
\cite[eq. (17)]{rsw}, we obtain images of $\rho_{1}$ with
$r=.546143, \theta=.752553, \phi=.351613$ and of $\rho_{2}$,
$r=.546138, \theta=.752544, \phi=.351621$. (Since 
the radial distance $r$ has not been 
preserved by $\Phi$, this  mapping  is not {\it unitary}.) 
The BH distance between these two
images is now $2.15078  \cdot 10^{-6}$, slightly 
{\it greater} than before the
application of $\Phi$.
If we {\it a priori} set $u=0$, so that $\Phi$ is {\it unital}, we found
we were more easily able to find such counterexamples to monotonicity. 
(For the benefit of the reader of \cite{rsw},
we point out that in the right-hand
side of eq. (2), the term  $w_{0}$ should occur as
a {\it common}
 factor and not just a factor of  $I$. Also there, the symbol $\bf{T}$ 
was meant --- as indicated to this author --- to refer only 
to the corresponding $4 \times 4$ matrix, and not to its 
$3 \times 3$ submatrix, where just $T$ was meant to be used. 
A correct presentation is given 
in \cite[sec. III]{ng}.)

On the other hand,  if one
lets the (maximum entropy) 
probability distributions corresponding
to $\rho_{1},\rho_{2}$ in the BH framework be $p_{1},p_{2}$,
then  $d_{Fisher}(T p_{1},T p_{2})
\leq d_{Fisher}(p_{1},p_{2})$, for {\it all} $p_{1}, p_{2}, T$.
We take $T$ to be 
 any (classical) stochastic mapping (Markov morphism),
and $d_{Fisher}$ to be the (classically unique) Fisher information
metric, given in our case by $ds^2_{BH}$. So, the inequality would be
violated for any strictly monotone (in the sense of
Petz and Sud\'ar \cite{ps}) metric.
\section{Comparative Noninformativities}
The {\it volume element} of the 
BH Fisher information metric (\ref{BHmetric}) 
is
\begin{equation} \label{BHve}
 { (r \coth({r \over 2})-2) 
\sqrt{4 -r^2 \mbox{csch}^{2}({r \over 2})} \sin{\theta} \over 4 r}.
\end{equation}
This can be normalized to a probability distribution (``Jeffreys'
 prior'', in Bayesian terminology) --- which we will denote by $p_{BH}$ --- over the
Bloch ball (unit ball in Euclidean 3-space) 
by dividing by its integral (.0983103) over the Bloch ball. If we
insert this particular integration 
value into the general formula for 
the asymptotic minimax/maximin redundancy for 
{\it universal} (classical) data compression
\cite[eq, (1.4)]{cb} (cf. \cite{kratt}),
\begin{equation} 
{d \over 2} \ln {n \over 2 \pi e} + \ln \int \sqrt{\det{I(\psi)}} 
\mbox{d} \psi + o(1),
\end{equation}
where $d$ corresponds to the dimensionality of the 
family of BH 
 probability distributions for which the Fisher information matrix 
$I(\psi)$ is being computed,
we get
\begin{equation}
 {3 \over 2}
\ln {n} -6.57644 + o(1).
\end{equation}
Here 
 $n$ is the number of observations and we take $d=3$ and $\sqrt{\det{I(\psi)}} = .0983103$.
For the universal {\it quantum} coding of two-level quantum
systems \cite{kratt} (cf. \cite{jozsa,hayashi,kalt}), on the other hand, 
the asymptotic minimax/maximin redundancy (which turns out, quite remarkably,
 to be intimately associated with the particular monotone metric
given by the {\it exponential} or {\it identric} mean 
$(1/e)(b^b/a^a)^{1/(b-a)}$ of numbers $a$ and $b$ --- the Bures metric 
itself simply being
associated with the  {\it arithmetic} mean $(a+b)/2$ \cite{ps}) is
\cite{GKS},
\begin{equation} 
{3 \over 2} \ln {n} - 1.77062 + o(1).
\end{equation}
The normalized form of the volume element of the 
(exponential/identric) {\it monotone} metric found in \cite{GKS} to
yield this quantum coding result is
\begin{equation}
p_{GKS} = .226321 (1-r)^{{1 \over 2 r} -1}
 (1+r)^{-{1 \over 2 r}-1} r^2 \sin{\theta}.
\end{equation}
(Another monotone metric that plays a distinguished role is the 
``Bogoliubov-Kubo-Mori'' one.  It has been shown by Grasselli and Streater
\cite{grasselli} that, in finite dimensions, 
this metric [and its constant multiples] is the only 
monotone one for which the (+1) and (-1) affine connections are mutually
dual. Also, the connection form [gauge field] pertaining to the generalization 
of the Berry phase to mixed states proposed by Uhlmann satisfies the
source-free Yang-Mills equation $*D*Dw=0$, where the Hodge star is
taken with respect to the Bures metric \cite{b3,me}.)

The BH Fisher information metric tensor {\it dominates} both the 
Bures (minimal monotone) and the {\it maximal} monotone (right logarithmic)
metric tensor over the entire Bloch ball, and therefore
all the (intermediate) 
monotone metric tensors. That is, if one subtracts
the tensor for any monotone metric from the BH one, the
eigenvalues of this difference are always nonnegative.

In \cite{slaterclarke} it was found that the normalized volume element
of the {\it Morozova-Chentsov} monotone metric,
\begin{equation}
p_{MC}= 0.00513299 (1-r^2)^{-{1 \over 2}} \log^{2}[{(1-r) \over (1+r)}] \sin{\theta},
\end{equation}
yielded a highly {\it noninformative} prior (a desideratum
in Bayesian analyses) in the class of monotone
metrics. By applying the original test of Clarke \cite{bernie} 
(implemented, in part, in \cite{slaterclarke}), we are 
now able to 
reach the conclusion that $p_{MC}$ is also considerably more noninformative
than $p_{BH}$.

 To develop our argument, 
let us denote the {\it relative entropy} of a probability distribution $p$
with respect to another such distribution $q$ by $D(p||q)$.
Then, we compute $D(p_{MC}||p_{BH})=1.99971$ and
$D(p_{BH}||p_{MC}) = 1.08908$. We regard $p_{MC}$, $p_{BH}$ as {\it prior}
distributions over the Bloch ball and convert them to {\it posterior} 
distributions $P_{MC}^{(3)},P_{BH}^{(3)}$ by multiplying them by 
the {\it likelihood}, $\Pi_{i=1}^{3} {(1-x_{i}^2) \over 2}$, 
that {\it three} pairs of measurements in the $x_{1},x_{2}$
and $x_{3}$ directions each yield one ``up'' and one ``down'' 
(the same 
likelihood function principally employed
 in \cite{slaterclarke};  cf. \cite[eq. (20)]{Bayesrule}). Normalizing the results, we are able to 
obtain that $D(P_{BH}^{(3)}||p_{MC})
=1.43453$ and $D(P_{MC}^{(3)}||p_{BH})= 1.67748$.
Since by {\it adding} information to $p_{MC}$ we render 
it {\it closer} to
$p_{BH}$ (that is, 1.67748 $<$1.99971), but {\it not} vice versa (1.43453 $>$ 
1.08908),
 the indicated conclusion is reached. These inequalities continue to hold 
 (although not so strongly) if instead
of imagining three pairs of measurements, we take {\it six}, two 
pairs in each direction, with
two ``ups'' and two ``downs'' resulting.
 Then, we have that $D(P_{BH}^{(6)}||p_{MC}) =
1.698$ (which is 
{\it still} less than 1.99971) and $D(P_{MC}^{(6)}||p_{BH})=1.74938$.
But this phenomenon stemming from the incorporation of 
 additional information into
$p_{MC}$ does not continue indefinitely, since
$D(P_{MC}^{(9)}||p_{BH}) = 2.02251$, which is now {\it larger}
than $D(p_{MC}||p_{BH}) = 1.99971$.
(We also computed here that $D(p_{MC}||p_{GKS})=.386051$, $D(p_{GKS}||p_{MC})=
.329118$ and $D(P_{MC}^{(3)}||p_{GKS}) = .188481$, $D(P_{GKS}^{(3)}||p_{MC})=
.771068$,
so $p_{MC}$ is more noninformative than $p_{GKS}$. )

Since the Morozova-Chentsov prior probability distribution was just
found to be considerably more noninformative than $p_{BH}$,
we investigated whether the {\it least} noninformative prior distribution
based upon a monotone metric \cite{slaterclarke},
 that is, the Bures (minimal monotone)
distribution,
\begin{equation}
p_{B} = {r^2 \sin{\theta} \over \pi^2 \sqrt{1-r^2}},
\end{equation}
was yet itself more noninformative than $p_{BH}$.
(Hall \cite{mjwhall} 
had interestingly 
 noted that the Bures metric over the Bloch ball of two-level
quantum systems was a specific form of the spatial part of 
the Robertson-Walker metric,
arising in general relativity \cite{weinberg}.)
This, in fact, 
 turned out to be the case, since $D(p_{BH}||p_{B}) = .221827$, 
$D(p_{B}||p_{BH})= .342287$, $D(P_{BH}^{(3)}||p_{B})=.432781$
and $D(P_{B}^{(3)}||p_{BH})=.2343$, which is {\it less} than
.342287. Also $D(P_{BH}^{(6)}||p_{B})=.662496$,
$D(P_{B}^{(6)}||p_{BH})= .306664$, which is still less than 
.342287, but $D(P_{B}^{(9)}||p_{BH})= .432335$.
So, the pattern is similar to that observed for $p_{MC}$, that is, the
improvement of noninformativity for the monotone metric prior over
$p_{BH}$
breaks down for {\it nine} pairs of measurements.
So, the posterior distributions based on {\it three} 
pairs of measurements in mutually orthogonal directions --- in both
the Morozova-Chentsov and Bures case --- give the best approximations,
in terms of relative entropy, to $p_{BH}$.

In Fig.~3 we show three {\it one}-dimensional {\it marginal} probability
distributions (obtained by integrating out $\theta,\phi$) 
over the radial coordinate $r$ of $p_{BH}$, $p_{B}$
and $P_{B}^{(3)}$. The marginal distribution of $p_{BH}$ is the 
most linear in nature of the three curves, while that for $p_{B}$ is the most
{\it steeply} ascending of the three, so it can be seen --- in line with our 
relative entropy
computations --- that the marginal
for the remaining curve, the posterior 
$P_{B}^{(3)}$,  is superior to the marginal for $p_{B}$
in approximating the marginal for the Brody-Hughston prior 
probability
distribution $p_{BH}$.
\newpage
\begin{figure}
\centerline{\psfig{figure=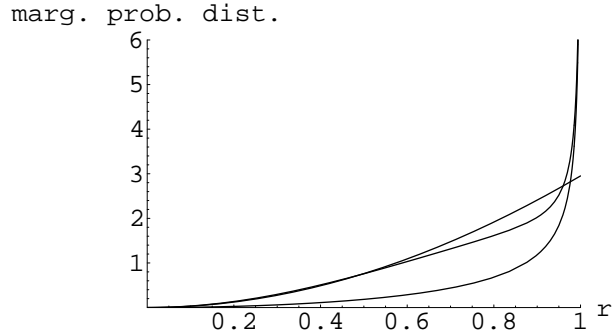}}
\caption{Marginal (one-dimensional) probability distributions over the radial
coordinate $r$ of $p_{BH}$ (the most linear of the three curves),
$p_{B}$ (the most steeply ascending) and 
the posterior $P^{(3)}_{B}$, which provides a superior 
 approximation to $p_{BH}$ than does  $p_{B}$.}
\end{figure}
In Fig.~4 we replace $P^{(3)}_{B}$ in Fig.~3 with
$P^{(3)}_{BH}$, thereby revealing graphically that adding
information to $p_{BH}$ makes it {\it less} resemble $p_{B}$,
as our computations in terms of relative entropy had indicated.
\begin{figure}
\centerline{\psfig{figure=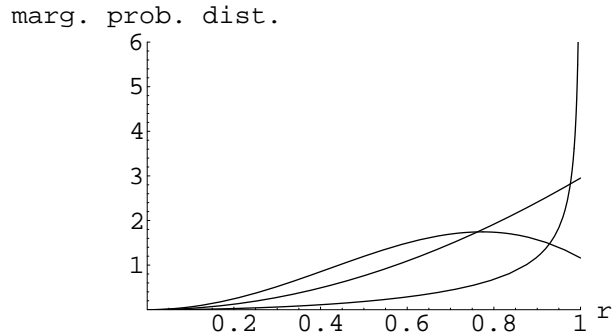}}
\caption{Same as Fig. 3, but for the replacement of $P^{(3)}_{B}$
by $P^{(3)}_{BH}$, which is {\it downward}-sloping for  $r>.7727551$ and {\it less}
resembles $p_{B}$ --- in terms of relative entropy --- 
than does $p_{BH}$ itself.}
\end{figure}
To continue further along these lines, we postulated a likelihood
based on {\it four} pairs of measurements, using the four diameters
of the Bloch sphere obtained from 
the vertices of an inscribed cube, each pair yielding
an ``up'' and a ``down''. Then, adopting our earlier notation,
we found that $D(P_{B}^{(4/\mbox{cube})}||p_{BH})= .0774351$, much {\it smaller}
than any of our previous relative entropy distances. In Fig.~5, we show
these two density functions.
\begin{figure}
\centerline{\psfig{figure=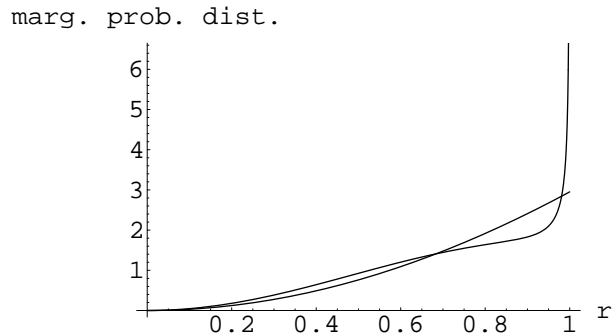}}
\caption{The (relatively flat) 
BH Fisher information 
distribution $p_{BH}$ along with our best approximation to it, 
the posterior distribution
$P_{B}^{(4/\mbox{cube})}$ formed from the Bures prior $p_{B}$ and
the likelihood that {\it four} pairs of 
spin measurements, each oriented along 
one of the four diameters
formed by an inscribed cube, will each yield an ``up''  and a ``down''}
\end{figure}
We then proceeded similarly, but now using {\it six} axes of 
spin measurement, based
on an inscribed (regular) {\it icosahedron}. We obtained 
a larger (that is, inferior) 
 figure of merit, $D(P_{B}^{(6/\mbox{icos})}||p_{BH})=.122255$.
\begin{figure}
\centerline{\psfig{figure=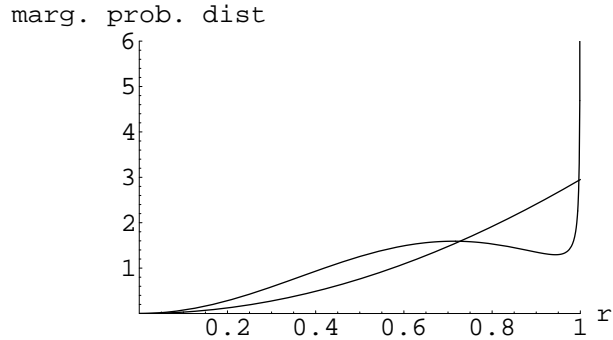}}
\caption{The posterior distribution $P_{B}^{(4/\mbox{cube})}$ in Fig.~5 is
replaced by (the more wiggly) $P_{B}^{(6/\mbox{icos})}$}
\end{figure}
With {\it ten} axes of spin 
measurement, based on an inscribed (regular) {\it dodecahedron},
we obtained a still larger relative entropy,
$D(P_{B}^{(10/dode)}||p_{BH})=.456816$.
\begin{figure}
\centerline{\psfig{figure=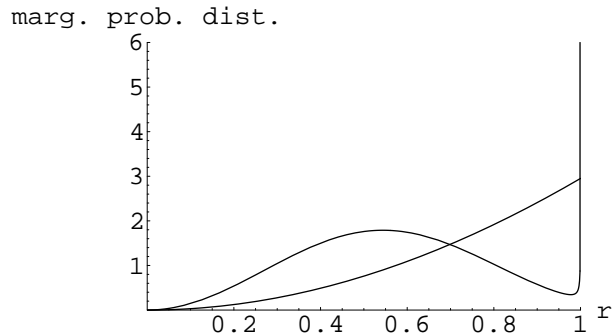}}
\caption{The posterior distribution 
$P_{B}^{(4/\mbox{cube})}$ in Fig.~5 is replaced by 
(the more wiggly) $P_{B}^{(10/\mbox{dode})}$}
\end{figure}
So, of the Platonic solids \cite{mike}, we have been able to approximate $p_{BH}$ most
closely with the use of measurements based on 
the {\it cube}. (Of course, our initial use
above and in \cite{slaterclarke} of {\it three}
mutually orthogonal 
axes of measurement corresponds to the use of an {\it octahedron}.) 
Exploratory efforts  of
ours employing  {\it non-separable} measurements 
\cite{slaterincreased} for the construction of
likelihood functions to use for converting the Bures prior probability
distribution $p_{B}$ to posteriors to well approximate $p_{BH}$, have not
so far been at all successful.

If we were to replace the radial coordinate of the metric $ds^2_{BH}$
by ${1\over 12 (1-r^2)}$, in an effort 
to render it {\it fully} monotone (and Fisher-adjusted) in nature,
then normalizing the resultant volume element to obtain the modified
probability distribution $p_{\tilde{BH}}$, we find that
 $D(p_{B}||p_{\tilde{BH}})= 8.36598 \cdot 10^{-6}$ and 
$D(p_{\tilde{BH}}||p_{B})= 8.37746 \cdot 10^{-6}$, that is both
much smaller relative entropies than previously obtained. (Actually, the
choice of the scaling constant 12 is irrelevant in this regard, since the
volume element just gets renormalized to the same probability
distribution in any case.)
Also, $D(P^{(3)}_{B}||p_{\tilde{BH}}) = .138763$ and
$D(P^{(3)}_{\tilde{BH}}||p_{B})= .143014$, so additional information --- as 
seems plausible --- does not diminish these 
two very small statistics.  We were not able --- using extended random 
searches --- to find
pairs of density matrices, the distance between the members of 
which increased under
stochastic mappings $\Phi$, thus not contradicting 
the possible monotonicity (we had sought to construct) of $d_{\tilde{BH}}$.
\section{Discussion}
It would be of interest to find and study 
the Brody-Hughston Fisher information
metrics for higher-dimensional quantum systems (such as
the {\it four}-dimensional {\it three}-level systems 
examined in \cite{slaterfour}) than the three-dimensional 
two-level ones
investigated here and in \cite{bh1}. However, our efforts along these lines
have yet to produce any simple, easily expressible results.

Another metric over the two-level quantum systems that seems of interest
to consider in the general 
context of this paper is \cite[eq. (16)]{slaterphysicaa},
\begin{equation} \label{BG}
ds_{BG}^{2} = 2 \Big({(1+r^2) \over (1-r^2)^2} dr^2 + {1 \over (1-r^2)} dn^2
\Big)
\end{equation}
This is the Fisher information 
metric that is obtained by adopting the point of view
of Bach \cite{bach1,bach2} and of Guiasu 
\cite{guiasu}, among others, that a density matrix can be considered as the
{\it covariance} matrix of a {\it complex multivariate normal distribution}
 over
the points of the corresponding Hilbert space.
(As with the approach of Brody and Hughston \cite{bh1},
 this too has a maximum-entropy
rationale, as a multivariate normal distribution has the maximum
entropy of all probability distributions having the same covariance matrix.)
If, as before, we equate the normal/tangential component to
$1/((1+r) f[(1-r)/(1+r)])$, we  obtain $f(t) = t/(1+t)$, which is
obviously a monotonically increasing function for $t>0$.
(In fact, the normal component of $ds_{BG}^{2}$ is simply twice that
for the {\it maximal} monotone metric, for which
$f(t) = 2 t/(1+t)$.)
Clearly, however, since the radial component of (\ref{BG}) is not
equal to $1/(1-r^2)$ (though its behavior [Fig. 8] 
for $r \in [0,1]$ is rather
similar), this metric is not strictly monotone in nature
(in the sense elaborated upon by Petz and Sud\'ar \cite{ps}),
though one might contend that it was ``approximatively'' monotone.
(At the outset of their paper \cite{zs}, Sommers and \.Zyczkowski
note that the {\it trace} metric on density matrices is monotone, but {\it 
not}
Riemannian, while the situation is reversed for the {\it Hilbert-Schmidt}
metric.)
\begin{figure}
\centerline{\psfig{figure=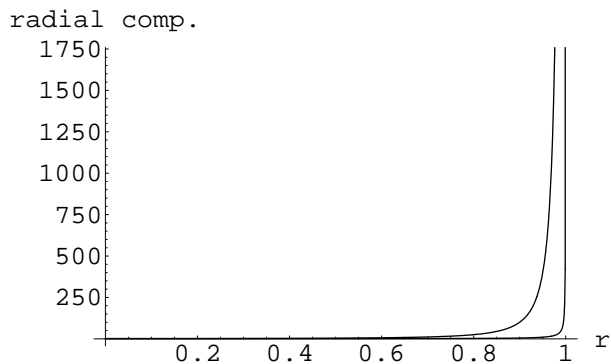}}
\caption{Radial components of an arbitrary monotone metric, that is, 
$1/(1-r^2)$, over
the Bloch ball and the (less-steeply ascending) radial component
of the Bach-Guiasu Fisher information metric (\ref{BG})}
\end{figure}

Brody and Hughston \cite{bh1} were interested in finding 
those probability
distributions over the pure states that yielded 
given density matrices and
were of maximum entropy.
Another analytical framework
 in which probability distributions
over the pure states arise
is in the quantum de Finetti Theorem as applied to density operators
with Bose-Einstein symmetry \cite{robin}.
But there, the probability distributions are {\it unique},
yielding exchangeable density operators,
so there is no recourse necessary to maximum entropy methods.
Let us 
further 
note that in their paper \cite{bh1}, ``Information content for quantum states,''
which has formed 
 the starting point for our analyses here,
Brody and Hughston sought the probability distribution over the pure
states that was ``{\it least informative} (their emphasis), subject to the
condition that it is consistent with the prescribed density matrix''.
It would seem somewhat paradoxical, then, at least at first glance,
that the normalized volume element (which we have denoted by $p_{BH}$) of 
the BH Fisher information metric for this  family of 
entropy-maximizing (information-minimizing) probability
distributions over the pure states, should itself be relatively
informative in nature (at least in comparison with the normalized
volume elements of the monotone metrics).
This phenomenon has been evidenced here by our
application of the 
interesting Bayesian methodology of Clarke \cite{slaterclarke,bernie}.

So, in conclusion, we have established here that if one adopts
the framework of Brody and Hughston, developed in \cite{bh1},
then one must sacrifice the desideratum of 
the exact monotonicity of metrics --- at
 least in the (quantum) sense of 
Petz and Sud\'ar \cite{ps}. Whether or not this 
``shortcoming'' comprises
a ``fatal flaw'' in the BH scheme
(as well as in the complex multivariate normal one 
 of Bach \cite{bach1,bach2} and Guiasu \cite{guiasu})
 appears to be a matter still open to some
discussion.

\acknowledgments

I would like to express appreciation to the 
Kavli Institute for Theoretical
Physics for computational  support in this research.

\end{document}